\documentclass[10pt,conference]{IEEEtran}

\usepackage{graphicx}
\usepackage{url}
\usepackage{amsfonts,amssymb,amsbsy}
\usepackage{amsmath}
\usepackage{tikz}

\usepackage{times}
\usepackage{tikz}
\usetikzlibrary{arrows,automata}

\usetikzlibrary{calc}
\usetikzlibrary{decorations.pathreplacing,decorations.markings,shapes.geometric}
\tikzset{naming/.style={align=center,font=\small}}
\tikzset{antenna/.style={insert path={-- coordinate (ant#1) ++(0,0.25) -- +(135:0.25) + (0,0) -- +(45:0.25)}}}
\tikzset{station/.style={naming,draw,shape=dart,shape border rotate=90, minimum width=10mm, minimum height=10mm,outer sep=0pt,inner sep=3pt}}

\definecolor{royalblue}{cmyk}{1,.5,0,0}
\definecolor{cerulean}{cmyk}{.94,.11,0,0}
\definecolor{violet}{cmyk}{.79,.88,0,0}
\definecolor {Eored}{rgb}{.647,.129,.149}
\definecolor {Eogreen}{rgb}{0,.53,0}

\usetikzlibrary{positioning} \definecolor{processblue}{cmyk}{.96,0,0,0}

\begin{document}

\title{Device-to-Device Data Storage for Mobile Cellular Systems}

\author{Joonas P\"a\"akk\"onen$^{\dagger}$, Camilla Hollanti$^{*}$, Olav Tirkkonen$^{\dagger}$
\\
$^\dagger$Department of Communications and Networking, School of Electrical Engineering, \\
$^*$Department of Mathematics and System Analysis, School of Science, \\
Aalto University, Espoo, Finland. \\
\{joonas.paakkonen, camilla.hollanti, olav.tirkkonen\}@aalto.fi}

\maketitle

\begin{abstract}
As an alternative to downloading content from a cellular access
network, mobile devices could be used to store data files and distribute them
through device-to-device (D2D) communication. We consider a D2D-based
storage community that is comprised of mobile users. Assuming that transmitting data from a
base station to a mobile user consumes more energy than transmitting
data between two mobile users, we show that it can be beneficial to
use redundant storage to ensure that data files stay available to the
community even if some of the storing users leave the network. We derive a
tractable closed-form equation stating when redundancy should be used
in order to minimize the expected energy consumption of data
retrieval. We find that replication is the preferred method of adding
redundancy as opposed to regenerating codes. Our
findings are verified by computer simulations.
\end{abstract}

\section{Introduction}\label{introductionsec}

The amount of mobile data traffic is growing tremendously. The total
global mobile traffic was about $885$ petabytes per month at the end
of 2012, and is expected to keep increasing \cite{cisco}. 
As traditional techniques for increasing the capacity of wireless
systems have their limits, new ways of reducing the load of the access
network are needed.

Recently, device-to-device (D2D) communication has been suggested as a
means of increasing the capacity and the throughput of cellular
systems, as well as improving the energy consumption of user devices,
see \cite{Kaufman}-\cite{belleschi}. As the storage capacity of
mobile devices increases, data files could be stored and retrieved from
the mobile users themselves in order to offload download traffic from the
infrastructure network. Distributed storage in Delay Tolerant, Ad Hoc
and D2D networks has been suggested
in~\cite{Ott2007,Lenders2007,golrezaei1}. In these, mobile terminals (with backup connections to an infrastructure network) are used to cache and distribute data files. To increase the
reliability of transmissions within the storage community, packet
level erasure coding is investigated in~\cite{pitkanen2007}.

In this paper, we concentrate on a system consisting of a base station and a
set of mobile users within the range of the base station, forming a
D2D storage community. The community consists of mobile users that will, sooner or later, leave the system. In order
to avoid losing stored data files, redundancy can be added to the stored data. The simplest way to do this is to store several copies of the files.
However, erasure coding can increase the performance of distributed data
storage \cite{compa}. Further, codes that are tailor-made for
distributed storage can improve system performance \cite{tailor}.

We apply regenerating codes \cite{dima} to a D2D
storage community and assess their performance. In~\cite{paakk}, we
investigated a similar system, under more complicated assumptions.
Here, we assume a wide-sense stationary storage community, with a
constant expected number of nodes. We assume that the community is able to recover and regenerate the lost data after each
single node departure before another departure takes place . We concentrate on the communication cost
incurred by file requests and storage regeneration, assuming that
the nodes have infinite storage capacities.

We find that, under the considered system assumptions, the simplest method of storing redundancy, i.e. storing one redundant replica of a file, is also the optimal method in terms of energy consumption.

It should be noted that, in this paper, we fully confine ourselves to assessing the theoretical performance of the storage and distribution methods at hand, and that we do not discuss the practical implementation of such methods. Likewise, we do not go into D2D device discovery, signaling, synchronization, power control, code construction etc.

The remainder of this paper is organized as follows: Section
\ref{systemmodelsec} explains the system model that we use throughout
this paper. Section \ref{analysissec} derives analytical expressions
for the select distribution methods. Section \ref{numericalresultssec}
shows both theoretical and simulated numerical results. Finally,
section \ref{conclusionssec} provides concluding remarks.

\section{System Model}\label{systemmodelsec}

We consider a wireless cellular system where mobile devices, referred
to as nodes, roam freely in and out of a geographically limited area.
We assume that the nodes themselves can be used to store (cache) data
and they can, upon request, transmit data to one another.

A set of nodes that are within a specified distance from each other
forms a storage community, or a local network. The local nodes can
communicate with each other in D2D mode, without the help of the
base station. Also, the base station can be used to transmit data to
the nodes but there is no need to relay data from a node to another
node via the base station.

\begin{figure}[tb]
\centering \includegraphics[scale=.3]{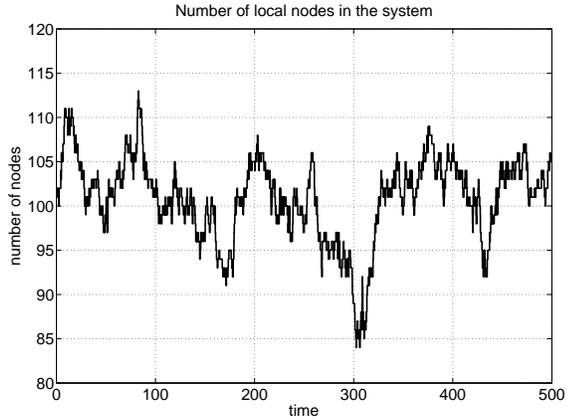}
\caption{An example realization of the number of nodes in the system. Here the expected number of nodes is $N=100$ and the expected node lifetime is $T=10$.}
\label{fluct}
\end{figure}

Nodes arrive in the system according to a Poisson process with
exponentially distributed inter-arrival times. The expected time for
which a single node sojourns in the system is denoted by $T$, the
expected \emph{node lifetime}. The expected \emph{number of nodes} in
the system is denoted by $N$. By Little's law \cite{little}, the
\emph{arrival rate} of the nodes is $\frac{N}{T}$. The expected
inter-arrival time of two consecutive nodes is $\frac{T}{N}$, which is also the
expected time between two consecutive node departures. These times are
exponentially distributed. The flow into system equals the flow out of
the system, and the number of nodes fluctuates around $N$. Fig. \ref{fluct} exemplifies this fluctuation.

\begin{figure}[b]
\begin{center}
\begin{tikzpicture}
[-latex,auto,auto,node distance=2cm and 2cm,on grid, semithick,state/.style={circle,thick,draw,blue, text=blue,minimum width=1.1cm}]
\node[state] (X) {$N$};
\node[state] (Xm) [left=of X] {$N{-}1$};
\node[state] (Xp) [right=of X] {$N{+}1$};
\node[state,draw=none] (finish) [left of=Xm] {...};
\node[state,draw=none] (end) [right of=Xp] {...};
\path (finish) edge [bend right=-45,color=Eogreen] node[above=0.1cm]{$N\lambda$} (Xm);
\path (Xm) edge [bend right=-45,color=Eogreen] node[above=0.1cm]{$N\lambda$} (X);
\path (X) edge [bend left=45,color=Eogreen] node[above=0.1cm]{$N\lambda$} (Xp);
\path (Xp) edge [bend right=-45,color=Eogreen] node[above=0.1cm]{$N\lambda$} (end);
\path (end) edge [bend right=-45,color=red] node[below=0.1cm]{$(N{+}2)\lambda$} (Xp);
\path (Xp) edge [bend right=-45,color=red] node[below=0.1cm]{$(N{+}1)\lambda$} (X);
\path (X) edge [bend left=45,color=red] node[below=0.1cm]{$N\lambda$} (Xm);
\path (Xm) edge [bend right=-45,color=red] node[below=0.1cm]{$(N{-}1)\lambda$} (finish);
\end{tikzpicture}
\end{center}
\caption{The M/M/$\infty$ Markov chain state diagram for the number of local nodes. The name of the state corresponds to the number of nodes (blue). The incoming rate (green) of the nodes is constant, whereas the outgoing rate (red) is proportional to the number of nodes in the system. The expected number of nodes is $N$ and $\lambda=1/T$.}
\label{mchain}
\end{figure}
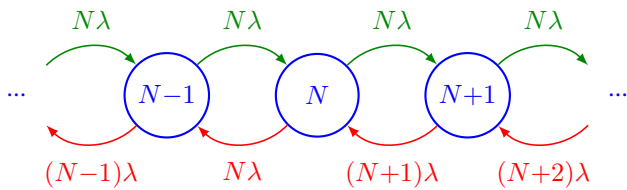

The time development of the number of local nodes can, thus, be
described with the M/M/$\infty$ Markov model, depicted in Fig.
\ref{mchain}. The steady-state probabilities for the M/M/$\infty$
model are well-known \cite{harrison}. The probability that there are
$i$ nodes in the system is
\begin{align}
\pi(i)=\frac{N^i}{i!}e^{-N}.
\label{poisson}
\end{align}

% THIS IS THE MARKOV CHAIN:

We assume that local nodes themselves can be used to cache
data. For simplicity, we assume that the storage capacity of each node
is infinite. We rationalize this by observing that the storage capacity of mobile devices has been dramatically increasing. This is why we presume that each node has some free
capacity that could be used for the common good.

The main motivation for assuming an infinite
storage capacity is that the storage problem of multiple files
decouples. Accordingly, it is sufficient to consider the storage
and distribution problem of a single file, with a specified request rate.

We denote the request rate of one file by one local node by $\omega$. The
inter-arrival time of two consecutive file requests follow the
exponential distribution with mean $\frac{1}{N\omega}$.

We normalize the size of the file to $1$ (bit). Similarly, we say that the cost (in
transmit energy) of transmitting one file from a local node to
another local node is also $1$ (joule). All the simplifying
assumptions mentioned here allow for tractable, tangible results.

\begin{figure}[htb]
\begin{tikzpicture}
\node[station] (base) {BS};
\draw[line join=bevel] (base.100) -- (base.80) -- (base.110) -- (base.70) -- (base.north west) -- (base.north east);
\draw[line join=bevel] (base.100) -- (base.70) (base.110) -- (base.north east);
\draw[line cap=rect] ([yshift=0pt]base.north) [antenna=1];
\foreach \x / \y in {{3.5/0.2/}}
\shade [ball color=royalblue] (\x,\y) circle (.3 cm);
\foreach \x / \y in {{5.0/0.5/},{6.3/-0.14/}}
\shade [ball color=Eogreen] (\x,\y) circle (.3 cm);
\foreach \x / \y in {{2.7/1.3/},{5.4/-1.3/},{3.7/-1.34/},{2.4/-0.66/},{4.85/1.7/},{6.0/1.25/},{7.0/-1.1/},{7.2/1.0/},{4.6/-0.6/},{4.1/1.0/}}
\shade [ball color=gray] (\x,\y) circle (.3 cm);
\path[thick] [-stealth,red,opacity=1] (0.5,1.2) edge node [sloped,below,text=black,opacity=1] {R} (3.0,0.4);
\path[thick] [-stealth,Eogreen,opacity=1] (4.6,0.4) edge node [sloped,below,text=black,opacity=1] {1} (3.9,0.25);
\end{tikzpicture}
\caption{A node (blue) requesting the file. Two local nodes (green) are caching a copy of the file (2-replication). Other nodes (gray) stay idle. We say that the cost of transmitting the file from a caching node is $1$, while the cost of transmitting the file from the base station (BS) is $R$.}
\label{BS}
\end{figure}
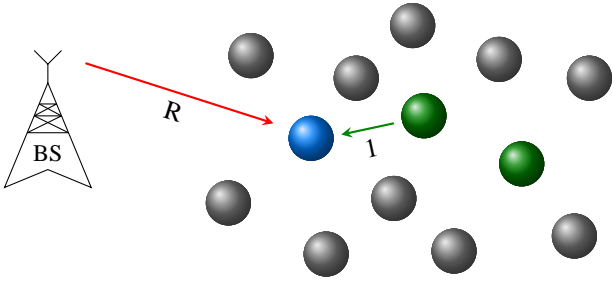

We assume that there is one data file. At random time instants, local nodes request the file and download it. The file can either be retrieved from the base station or from the
local nodes through D2D communications. It is, on average, $R$ times as
expensive to download a bit from the base station as compared to
downloading a bit from another local node, with $R>1$. The caching
model is depicted in Fig. \ref{BS}. 

The downloading node can download the file from the local nodes only
if the file is cached. In this paper, we compare two caching methods:
\begin{itemize}
\item \emph{Simple caching}: If the requested file is already cached on another local node, the caching node transmits the file to the requesting node in D2D mode. If the file is not cached on any of the local nodes, the base station transmits the file to the requesting node. Thence, the requesting node caches the file and, later on, transmits it to other users upon request. Only one local node at a time is caching the data file and, thus, there is no redundancy.
\item \emph{Redundant caching}: A subset of the local nodes is used to transmit parts of the file to the downloading node and the original file is reconstructed at the downloading node. Two or more nodes are caching the file or a fraction of the file. One of the caching nodes is redundant.
\end{itemize}
The simplest way of redundant caching is allocating two exact replicas
of the whole file on two different nodes. We call this method
\emph{2-replication}.

Retrieving a file from the base station is never beneficial as long as
the file is available in the storage community, and it is more
expensive to retrieve data from the base station than to retrieve data from another
node. 

We assume that the file is always available -- only the cost (in
transmit energy) and the data traffic load on the base station change
depending on the distribution method. Whether it is beneficial to use
\emph{simple caching} or \emph{redundant caching} depends on the
system parameters and the popularity (request rate) of the requested file.

% THIS IS THE BASE STATION:
We define the \emph{cost} as the expected total amount of transmit energy per time unit that must be used by the local nodes and the base station. Our objective is to find expressions for the expected total cost of different distribution methods given the system parameters $R, N, \omega$ and $T$. Eventually, we find the distribution method that yields the smallest expected cost given the aforementioned system parameters.

\section{Analysis}\label{analysissec}

In this section, we derive closed-form expressions (approximations)
for the expected total costs of simple caching and redundant caching.
Later in this section, we compare these methods with each other.

\subsection{Simple caching}

Initially, suppose that the file is already cached on one of the local
nodes. Thus, as long as the node that is caching the file stays in the
system, all file requests result in retrievals from this node. There
are on average $N$ local nodes that generate requests, each at rate
$\omega$, and the expected lifetime of any of the nodes is $T$.
Therefore, the expected number of requests during the lifetime of the
caching node is $N\omega T$.

Now suppose that the cost of retrieving the file (of size $1$) from
another local node is simply $1$. Hence, the expected cost of
downloading the file from the base station is $R$. If the caching node
has left the system, the next node that requests the file has to
download it from the base station. The expected time in which this
happens is $\frac{1}{N\omega}$ as the expected total request rate is
$N\omega$. Thus, the time in which an expected number of $N\omega
T{+}1$ requests are generated is $T{+}\frac{1}{N\omega}$. The expected
cost of these requests is $N\omega T{+}R$ and, thereby, the expected
cost of simple caching becomes:
\begin{align}
C_{\text{sc}}(R,N,\omega,T) = \frac{N\omega T + R}{T + \frac{1}{N\omega}} = \frac{N^2\omega^2 T + RN\omega}{1 + N\omega T}.
\label{simp}
\end{align}

\subsection{Redundant caching}
Here we use a $(n,k,d)=(k{+}1,k,k)$ regenerating code \cite{dima} to
cache the file on a set of local nodes in a distributed manner. Thus,
any $k$ nodes that are caching an encoded fraction of the file can be
used to reconstruct or repair the file. The file is fractioned into
$k$ encoded blocks and one block is allocated to $k{+}1$ different
caching nodes. One block is redundancy, and $k=1,2,3,...$. Hence,
should any of the caching nodes leave the system, the remaining
(\emph{surviving}) $k$ nodes can be used to \emph{regenerate} the lost
block.

\begin{figure}[tb]
\centering \includegraphics[scale=.23]{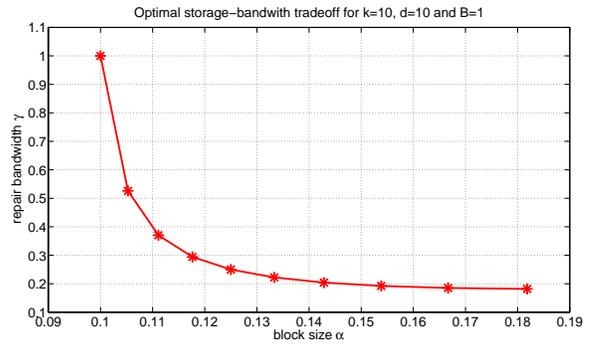}
\caption{Regenerating codes can be used to repair a lost encoded block by only transmitting a number of data equal to the block size. The code that achieves this property is called the Minimum Bandwidth Regenerating (MBR) code (rightmost point). Traditional erasure coding (leftmost point) requires the whole data object to be communicated. Here the file size ($B$) is $1$.}
\label{tradeoff}
\end{figure}

The repair bandwidth of a regenerating code is defined as the number of data communicated
when a lost block is regenerated. As we consider infinite storage capacities, only the repair bandwidth of is relevant. For this reason, we choose to use the
Minimum Bandwidth Regenerating (MBR) code (see Fig. \ref{tradeoff}).

Whenever there is a \emph{failure}, i.e. one of the caching nodes
leaves the system, the lost block is repaired to another local node.
This requires $\gamma(k) = \frac{2}{k+1}$ bits to be transmitted
for the MBR code with repair degree $d=k$ \cite{dima}. The repair
bandwidth of an MBR code is equal to the size of the encoded (cached)
block $\alpha(k)$. Thus, in total, $k\alpha(k) = k\gamma(k) =
\frac{2k}{k+1}$ bits must be transmitted whenever a local node
downloads the file from a set of $k$ caching nodes.

\begin{figure}[htb]
\centering
\begin{tikzpicture}
\foreach \x / \y in {{3.5/0.2/}}
\shade [ball color=yellow] (\x,\y) circle (.3 cm);
\foreach \x / \y in {{6.3/-0.14/}}
\shade [ball color=orange] (\x,\y) circle (.3 cm);
\foreach \x / \y in {{5.0/0.5/}}
\shade [ball color=Eogreen] (\x,\y) circle (.3 cm);
\foreach \x / \y in {{2.7/1.3/},{5.4/-1.3/},{3.7/-1.34/},{2.4/-0.66/},{4.85/1.7/},{6.0/1.25/},{7.0/-1.1/},{7.2/1.0/},{4.6/-0.6/},{4.1/1.0/}}
\shade [ball color=gray] (\x,\y) circle (.3 cm);
\path[thick] [-stealth,red,opacity=1] (6.6,-0.14) edge node [sloped,below,text=black,opacity=1] {departure} (8.9,-0.14);
\path[thick] [-stealth,Eogreen,opacity=1] (4.6,0.4) edge node [sloped,below,text=black,opacity=1] {repair} (3.9,0.25);
\end{tikzpicture}
\caption{Repair of 2-replication. If a caching node (orange) leaves the system, the surviving caching node (green) can repair the file by sending a copy of the file to an idle node. This node stores the copy and, thereby, becomes the new caching node (\emph{newcomer}, yellow).}
\label{failure}
\end{figure}
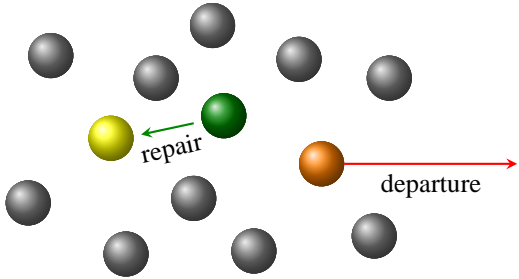

Next, we derive an approximation for the expected cost of redundant
caching with the $(k{+}1,k,k)$ regenerating code. We note that the
expected state of the system is such that there are $N$ nodes in the
system and $k{+}1$ out of these $N$ nodes are caching an encoded data
block. The expected sojourn time of all of these $N$ nodes in the system
equals the expected node lifetime $T$. When one of these caching nodes
leaves, $\gamma(k) = \frac{2}{k+1}$ bits need to be communicated in
order to repair the lost data block and store it on another node. Setting the cost of transmitting a bit from a local node to another local node
to $1$, the expected repair cost becomes
\begin{align}
C_{\text{x}}(k,T) = \frac{k+1}{T} \times \gamma(k) = \frac{k+1}{T} \frac{2}{k+1} = \frac{2}{T},
\label{repaireq}
\end{align}
which is, interestingly, independent of $k$, and equals the cost of
the repair process of 2-replication\footnote{The expected repair cost of 2-replication is $\frac{2}{T} \times 1$ as there are two blocks, each of size $1$ (the file size).} 
The process of 2-replication is depicted in Fig. \ref{failure}. Even
though increasing $k$ decreases the repair bandwidth $\gamma$, it also
increases the expected number of failures, as a failure takes place
whenever a caching node leaves the system. These effects cancel out each other.

For simplicity, let us assume that the number of local nodes never drops below $k$ and the repair process is so fast (immediate) that no nodes leave the system before the repair process is complete. Thence, we never need to reallocate data from the base station, and only repairs incur upkeep costs. Therefore, the expected cost per time unit of redundant caching becomes
\begin{align}
C_{\text{rc}}(k,N,\omega,T) = N\omega k \alpha(k)+ C_{\text{x}}(k,T) = N\omega \frac{2k}{k+1} + \frac{2}{T}
\nonumber
\end{align}
as all requests result in local downloads and the expected cost of retrieving (reconstructing) the file is $\frac{2k}{k+1}$. It is easy to see that $C_{\text{rc}}$ is minimized at $k=1$. This is not a regenerating code -- \emph{the method that minimizes the expected total cost of redundant caching is 2-replication}. Note that $k{=}1$ also minimizes the probability that a file request results in a local download. This is because in order for a local request to take place, there must to be at least $k$ nodes in the system. The expected cost of 2-replication becomes
\begin{align}
C_{\text{2-rep}}(N,\omega,T) = N\omega + \frac{2}{T}.
\label{rep}
\end{align}
It should be noted that more than two copies of the file could be replicated on the nodes. The derivation of the cost in this case would be similar. Having more than just two copies of the file would enable the system to withstand more than one caching node leaving the system. However, as we assume that the file can be repaired before another node leaves the system, we restrict ourselves to the case where there is only one redundant copy of the data file in the system.

Besides having the smallest possible repair bandwidth, another benefit of 2-replication over regenerating codes is its simplicity. There is no need to perform excessive computations when the file is reconstructed or requested---the file is simply copied from a caching to the requesting node. Similarly, at repair, the file is simply copied to the newcomer node. See Fig. \ref{failure} for an illustration of the repair process.

% THIS IS THE BUNCH OF NODES AT REPAIR:

If 2-replication is used, the file, or a redundant copy of the file,
needs to be reallocated from the base station only if the number of
nodes drops below two. According to (\ref{poisson}), the probability
of this is $\frac{N+1}{e^N}$. For large $N$, we can approximate this
to be zero (for instance, already for $N=20, \frac{21}{e^{20}} \approx
4.33\times 10^{-8}$). This is the reason why we ignore the cost of
reallocating the file to the nodes from the base station. This is also
the reason why we assume that, when 2-replication is used, there is
always a node to which we can copy the file whenever a caching node
leaves the system. This allows us to approximate the total repair cost
in~(\ref{repaireq}) as $\frac{2}{T}$, as discussed earlier.

\subsection{Comparison}
Here we derive a straightforward decision rule on when to use simple
caching (without redundancy) and when to use redundant caching
(2-replication with one redundant copy). Simply by setting
$C_{\text{sc}} > C_{\text{2-rep}}$ (\eqref{simp}, \eqref{rep}), we
find that redundant caching outperforms simple caching if
\begin{align}\nonumber
\frac{N\omega T + R}{T + \frac{1}{N\omega}} > N\omega + \frac{2}{T},
\end{align}
which yields
\begin{align}
R > 3 + \frac{2}{N\omega T}.
\label{R3region}
\end{align}

\begin{figure}[tbhp]
\centering \includegraphics[scale=.3]{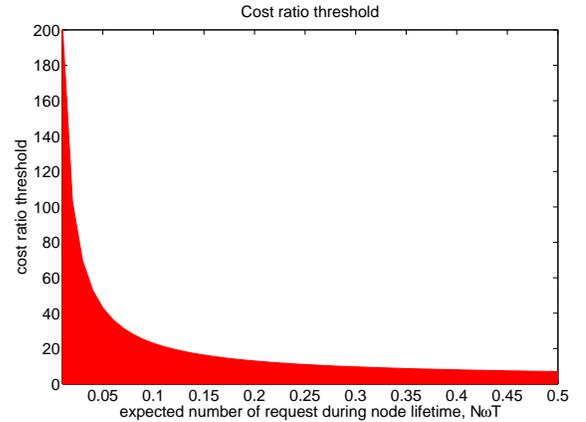}
\caption{Cost ratio threshold. Whenever $R>3+\frac{2}{N\omega T}$ (white region), 2-replication yields the lowest expected cost. Otherwise (red region), simple caching should be used instead.}
\label{region}
\end{figure}

Fig. \ref{region} shows the decision boundary of \eqref{R3region}.
It is interesting to note that as long as $R\leq 3$, the best method
is, independently of the other parameters, simple caching. Also, note
that $N\omega T$ can be interpreted as the expected number of file
requests made in the system during the lifetime of a single node. For example, if the expected number of requests during the lifetime of a
node is greater than two, $R\geq 4$ is enough to justify redundant
caching.

It might seem tempting to use redundant caching (2-replication) over
simple caching whenever condition \eqref{R3region} is met. However,
simple caching only takes up half of the storage space of
2-replication. Consequently, simple caching can store twice as many
files as 2-replication. In addition, 2-replication requires a D2D
connection to be established for the repair process whenever a caching
node leaves the system.

Even though D2D data distribution may reduce the traffic load on the base station and decrease the overall power consumption, it should be noted that the power consumption of the users that store and distribute data may increase considerably. This is why the caching users should be provided with perks, e.g. they could be granted more download bandwidth.

Table \ref{tab} concludes this section by comparing the expected costs (or their approximations) of the considered caching methods.
\begin{table}[htb]
\begin{center}
\caption{Comparison of caching methods}
\def\arraystretch{2}
\begin{tabular}{ | c | c | c |}
\hline
\textbf{method} & \textbf{caching nodes} & \textbf{cost (per time)}  \\ \hline
base station only & $0$ & $RN\omega$ \\ \hline
simple caching & $1$ & $\frac{N^2\omega^2 T + RN\omega}{1 + N\omega T}$ \\ \hline
2-replication & $2$ & $\approx N\omega + \frac{2}{T}$ \\ \hline
regenerating code & $k+1\geq 3$ & $\approx N\omega \frac{2k}{k+1} + \frac{2}{T}$ \\ \hline
\end{tabular}
\label{tab}
\end{center}
\end{table}

\newpage
\section{Numerical Results}\label{numericalresultssec}
This section provides simulation results of the expected cost for simple caching and 2-replication. All the simulations are conducted over $2000 T$ time units, where $T$ is the expected node lifetime, and the average cost per time unit over the runs is presented. Simulation results are compared with the theoretical results. Overall, it can be concluded that the expected theoretical values coincide with the average simulated values. However, there is some (yet negligible) discrepancy due to the random nature of the simulations.

\begin{figure}[tbhp]
\centering \includegraphics[scale=.3]{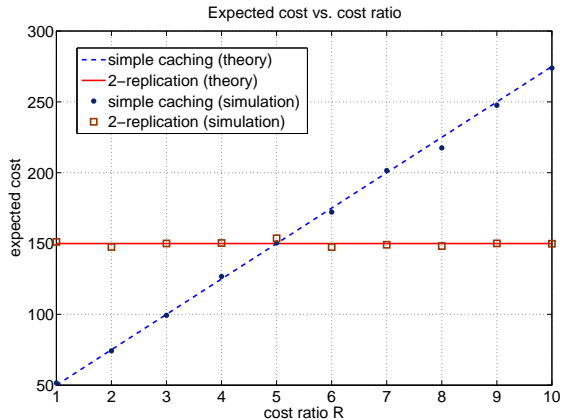}
\caption{Expected cost vs. cost ratio $R$ with parameter values $N=100, \omega=0.5$ and $T=0.02$. The cost of simple caching is linear w.r.t. $R$, while the cost of 2-replication is practically independent of $R$. The simulation results are well in line with Equation \eqref{R3region}; 2-replication outperforms simple caching as long as $R > 3 + \frac{2}{N\omega T} = 5$.}
\label{costvR}
\end{figure}

\begin{figure}[tb]
\centering \includegraphics[scale=.3]{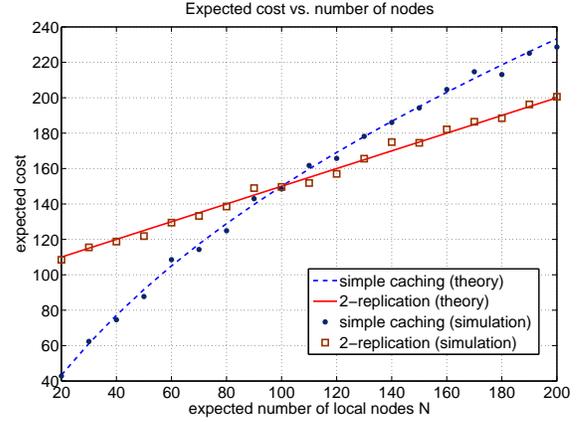}
\caption{Expected cost vs. expected number of nodes $N$ with parameter values $R=5, \omega=0.5$ and $T=0.02$.}
\label{costvN}
\end{figure}

Figures \ref{costvR} and \ref{costvN} 
illustrate the expected theoretical costs and the average simulated costs as
functions of the expected cost ratio $R$ and the expected number of nodes
$N$, respectively. The expected
cost behaves similarly as a function $N$ and $\omega$ (see Table
\ref{tab}). 

\begin{figure}[tb]
\centering \includegraphics[scale=.3]{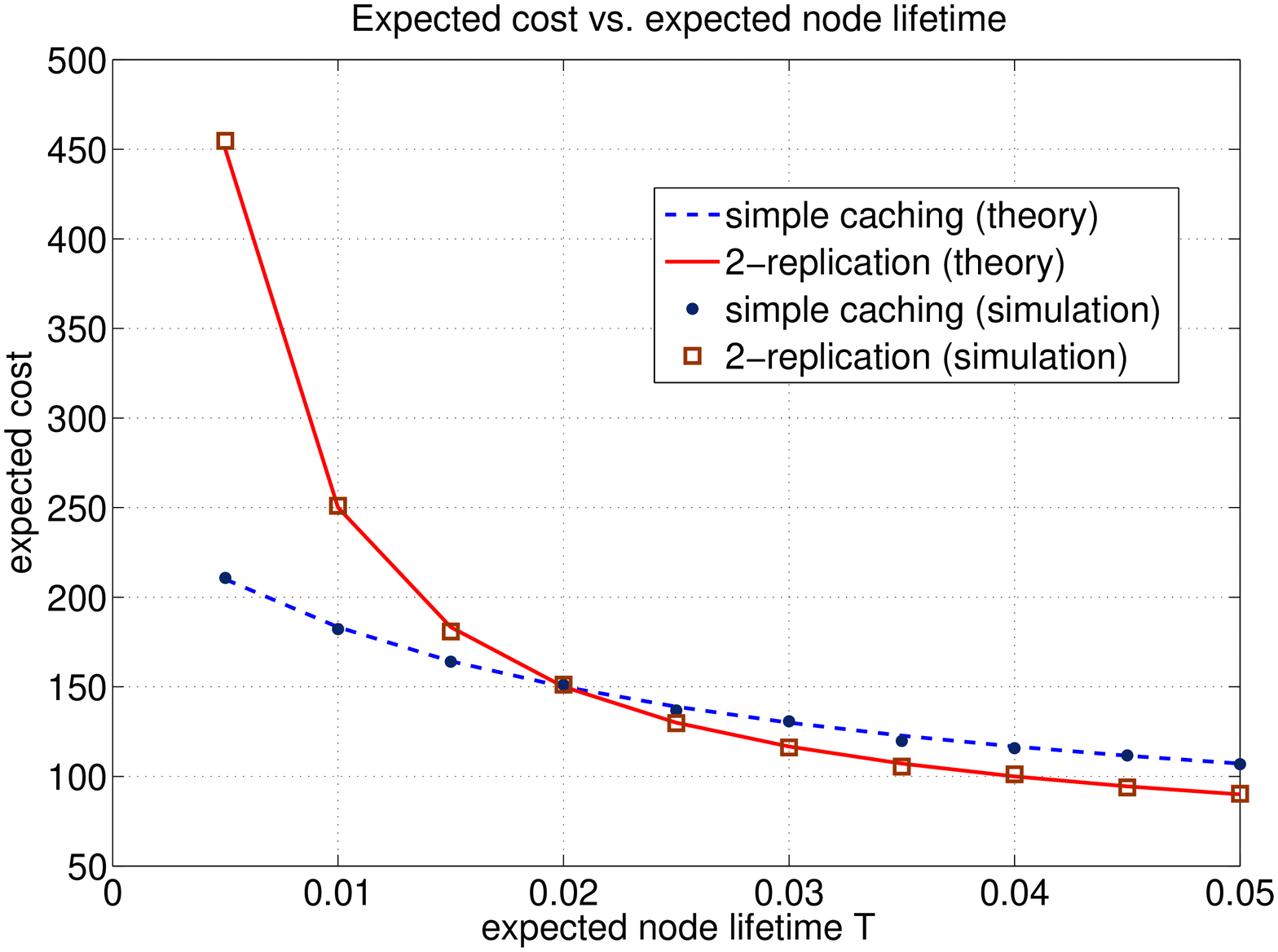}
\caption{Expected cost vs. expected node lifetime $T$ with parameter values $R= 5, N=100$ and $\omega=0.5$. As $T$ tends to infinity, the expected cost of simple caching tends to $N\omega=50$ as does that of 2-replication (Table \ref{tab}). As $T$ tends to $0$, the expected cost of 2-replication tends to infinity, while the expected cost of simple caching tends to $RN\omega = 250$ \eqref{simp}.}
\label{costvT}
\end{figure}

Finally, Fig. \ref{costvT} shows the expected theoretical costs and the average simulated costs as functions of the expected node lifetime $T$. As $T$ tends to infinity, the expected cost of simple caching tends to that of 2-replication, namely, $N\omega$ (Table \ref{tab}). This means that if the nodes stay in the system for a long period of time, all the file requests result in local downloads from the caching nodes and the distribution method is irrelevant. Conversely, if $T$ tends to $0$, the expected cost of 2-replication tends to infinity, while the expected cost of simple caching tends to $RN\omega$ \eqref{simp}. Thus, 2-replication should not be used for highly unstable systems with short node lifetimes -- a short node lifetime implies a high departure rate of caching nodes and, consequently, a high repair cost.

\section{Conclusions}\label{conclusionssec}
We have shown that, for the $(k{+}1,k,k)$ regenerating code, the expected total repair bandwidth is practically independent of $k$ and coincides with that of 2-replication. Also, we have demonstrated that, under our assumptions, the expected total cost of 2-replication is lower than that of the aforementioned regenerating code. Finally, we have found a simple decision rule for choosing between simple caching and 2-replication in order to minimize the expected total cost in terms of energy consumption.


\begin{thebibliography}{1}

\bibitem{cisco}
Cisco Visual Networking Index: Global Mobile Data Traffic Forecast Update, 2012-2017. [Online]. Available: http://www.cisco.com/en/US/solutions/collateral/ns341/ns525/ns537/\\ns705/ns827/white\_paper\_c11-520862pdf.

\bibitem{Kaufman} B. Kaufman, and B. Aazhang, "Cellular networks with an overlaid device to device network," \emph{Asilomar Conference on Signals, Systems and Computers}, pp. 1537-1541, Oct. 2008.

\bibitem{chia}
C.-H. Yu, O. Tirkkonen, K. Doppler and C. Ribeiro, ``On the
performance of Device-to-Device underlay communication with simple
power control,'' \emph{IEEE VTC--Spring}, 5 pp., April 2009.

\bibitem{belleschi}
M. Belleschi, G. Fodor, and A. Abrardo, ``Performance Analysis of a Distributed Resource Allocation Scheme for D2D Communications,"
\emph{IEEE Workshop on Machine-to-Machine Communications}, December 2011.

\bibitem{Ott2007} J. Ott and M. Pitk\"anen. DTN-based Content Storage
  and Retrieval. \emph{IEEE WoWMoM Workshop on Autonomic and
    Opportunistic Communications (AOC)}, June 2007.

\bibitem{Lenders2007} V. Lenders, G. Karlsson, and M. May, ``Wireless
  Ad Hoc Podcasting,'' \emph{IEEE Conference on Sensor, Mesh, and Ad
    Hoc Communications and Networks (SECON)}, June 2007.

\bibitem{golrezaei1}
N. Golrezaei, A. G. Dimakis, and A. F. Molisch, ``Base Station Assisted Device-to-Device Communications for High-Throughput Wireless Video Networks,"
\emph{IEEE International Conference on Communications, Realizing
  Advanced Video Optimized Wireless Networks}, 2012.

\bibitem{pitkanen2007} M. Pitk\"anen, and J. Ott, "Redundancy and distributed
  caching in mobile DTNs," \emph{CM/IEEE international workshop on Mobility in the evolving internet architecture}, 7 pp. Aug. 2007. 

\bibitem{compa}
H. Weatherspoon, and J. D. Kubiatowicz, ``Erasure coding vs. replication: a quantitative comparison,"
in \emph{Proc. International Workshop on Peer-to-Peer Systems}, March, 2002.

\bibitem{tailor}
A. Datta, and F. Oggier, ``An Overview of Codes Tailor-made for Networked Distributed Data Storage,"
\emph{Association for Computing Machinery Special Interest Group on Algorithms and Computation Theory News}, vol. 44, no. 1, March 2013, pp. 89-105.

\bibitem{dima}
A. G. Dimakis, P. B. Godfrey, Y. Wu, M. O. Wainwright, and K. Ramchandran, ``Network coding for distributed storage systems,"
\emph{IEEE Transactions on Information Theory}, vol. 56, no. 9, pp. 4539-4551, September, 2010.

\bibitem{paakk}
J. P\"a\"akk\"onen, P. Dharmawansa, C. Hollanti, and O. Tirkkonen, ``Distributed Storage for Proximity Based Services,"
in \emph{Proc. IEEE Swedish Communication Technologies Workshop}, October, 2012, pp. 30-35.

\bibitem{little}
Allen, A. O., \emph{Probability, Statistics, and Queueing Theory: With Computer Science Applications.} Gulf Professional Publishing, 1990, pp. 259.

\bibitem{harrison}
P. Harrison, and N. M. Patel, \emph{Performance Modelling of Communication Networks and Computer Architectures.} Addison-Wesley, 1992, pp. 173.

\end{thebibliography}
\end{document}